# Photon momentum enabled light absorption in bulk silicon

Sergey S. Kharintsev[1*], Aleksey I. Noskov[1], Elina I. Battalova[1], Liat Katrivas[2], Alexander B. Kotlyar[2], Jovany Merham[3], Eric O. Potma[3], Vartkess A. Apkarian[3], and Dmitry A. Fishman[3*]

[1]Department of Optics and Nanophotonics, Institute of Physics, Kazan Federal University, Kazan 420008, Russia

[2]George S. Wise Faculty of Life Sciences, Tel Aviv University, Tel Aviv 6997801, Israel

[3]Department of Chemistry, University of California Irvine, Irvine, CA 92697, USA

\* skharint@gmail.com

   dmitryf@uci.edu

ABSTRACT

Photons do not carry sufficient momentum to induce indirect optical transitions in semiconducting materials such as silicon, necessitating the assistance of lattice phonons to conserve momentum. Compared to direct bandgap semiconductors, this renders silicon a less attractive material for a wide variety of optoelectronic applications. In this work, we introduce an alternative strategy to fulfill the momentum-matching requirement in indirect optical transitions. We demonstrate that when confined to scales below ~3 nm, photons acquire sufficient momentum to allow electronic transitions at the band edge of Si without the assistance of a phonon. Confined photons allow simultaneous energy and momentum conservation in two-body photon-electron scattering; in effect, converting silicon into a direct bandgap semiconductor. We show that this less-explored concept of light-matter interaction leads to a marked increase in the absorptivity of Si from the UV to the near-IR. The strategy provides opportunities for more efficient use of indirect semiconductors in photovoltaics, energy conversion, light detection and emission.

INTRODUCTION

Both energy and momentum must be conserved in light-matter interactions. For optical transitions in semiconducting materials, this requirement dictates that the energy difference between material states must match the photon energy. Because the photon momentum in free space is negligible compared to the momenta of electrons, satisfying the two conservation laws is conditional on the vertical alignment of the bottom of the conduction band and the top of the valence band. When aligned, the bandgap is classified as direct and the transition is optically allowed, as depicted by vertical arrows in the material's energy band diagram. In case the states are offset in *k*-space, the bandgap is indirect. This scenario applies to the bandgap of silicon at 1.1 eV (at $\lambda_0$=1130 nm). As can be seen by the band structure provided in Fig. 1a, the band extrema at the $\Gamma$ and $X$ points are separated by a lattice momentum of $\sim 2\pi\hbar/a$, where $a = 0.54$ nm is the lattice constant, which is three orders of magnitude ($\lambda_0/a$) larger than the momentum of a 1.1 eV free space photon. While direct optical transitions at this energy are forbidden, momentum conservation can still be satisfied through indirect, phonon-assisted transitions. However, the three-body electron-photon-phonon scattering nature of indirect transitions is responsible for the two orders of magnitude weaker absorptivity of Si near its band edge, relative to direct bandgap semiconductors [1,2], as illustrated in Fig. 1c.

Despite silicon's preeminence in optoelectronics, its unfavorable optical properties impose fundamental limitations on many optoelectronic technologies. Such limitations have implications for solar energy conversion and photovoltaic industries, where Si strongly dominates[3]. Because of silicon's modest light absorption, over 90% of currently installed solar cells use crystalline silicon with a thickness of more than 100 µm to achieve efficient light-to-energy conversion throughout the visible and near-infrared range (Supplementary Figure SF1). In contrast, in direct bandgap semiconductors, similar conversion efficiencies can be reached at thicknesses of 0.2 µm due to their greater absorptivity[4]. Since the price of conventional solar cells is intimately related to the cost of the semiconductor material, silicon's weak absorption coefficient lowers its practical implementation as a competitive renewable energy resource [9], and challenges its promise as a viable alternative to fossil fuels [10].

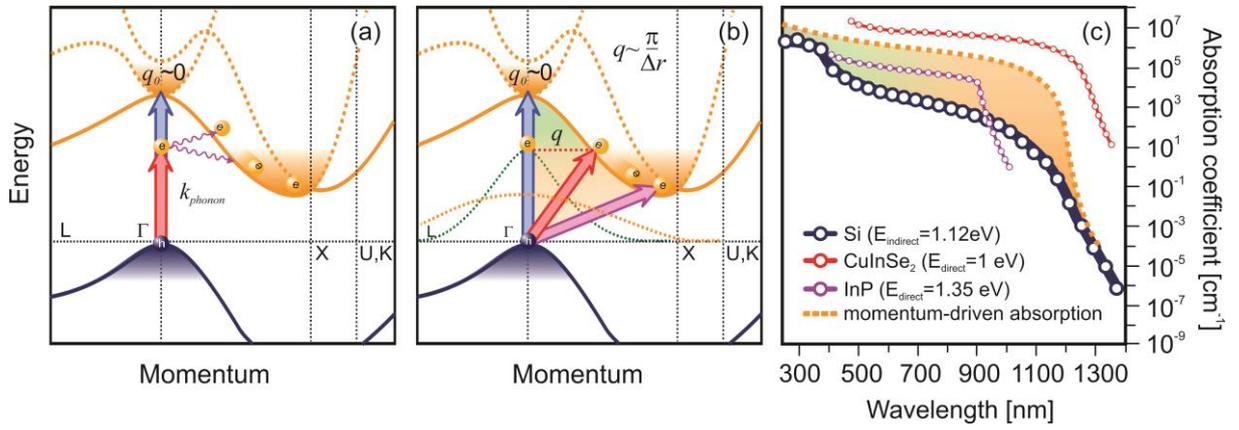

**Figure 1.** (a) Energy-momentum diagram for optical transitions in pure Si with free space photons. The arrows indicate direct (blue arrow, $q_0 \sim 0$) and indirect optical transitions (red and velvet arrows, $k_{phonon} > 0$). (b) Energy-momentum diagram showing optical transitions enabled by the momenta of confined photons, without the need of lattice phonons. The momenta distributions are schematically depicted for to spatial confinement $\Delta r \sim 0.7$ nm (orange curve) and $\Delta r \sim 1.4$ nm (green curve). Both diagrams (a) and (b) depict optical transitions near the Γ point, excluding other momentum-enabled transitions at various other points across the Brillouin zone. (c) Absorption spectrum of Si [5], and direct bandgap semiconductors InP [6,7] and $CuInSe_2$ [8]. Dashed curve and shaded region indicate the increase of the absorption coefficient across a broad spectral range due to photon momentum-enabled transitions.

This challenge has spurred numerous efforts to improve light management in silicon [11]. For instance, light trapping in black[12] or textured silicon[13] has been implemented for enhanced absorption. Other efforts employ subwavelength structures, such as nanocones [14], to trap photons by increasing the effective refractive index or the local density of optical states [15]. Such approaches include the use of plasmonics, typically in the form of metallic nanoparticles that act as near-field concentrators[16]. Yet, the performance of plasmonic structures is limited by scattering and ohmic losses[17]. Moreover, plasmonic resonances are typically narrow in comparison to the solar spectrum, no part of which can be spared.

While the majority of the efforts aim at increasing the photon density that reaches a weak absorber, we pursue a fundamentally different strategy. We depart from the tenet that absorption strength is solely a material property. Optical transitions are determined by the light-matter interaction, therefore the transition probability is as much a property of the employed light as the material (see *Supplementary Information Part 2*). The photon momentum can be boosted by 2-3 orders of magnitude to match the lattice momentum of Si, by confining it on nanometric scales, $\Delta r$. Consequently, it can be expected that Si effectively behaves as a direct bandgap semiconductor when using confined photons whose momentum distribution is $\Delta p \sim \pi \hbar / \Delta r$ [18,19]. Photons can be most tightly confined in plasmonic cavities or junctions, where local fields are carved by the atomistic morphology of the structure [20]. Such picocavities [21,22,23] can reach effective volumes of $V_e < (1\ nm)^3$. Confinement of photons at this length scale has been visualized with tip-enhanced Raman microscopy, revealing a Gaussian-like light field intensity distribution of standard deviation $\sigma_x = 0.13$ nm (0.16 nm image resolution) [24,25].

The first Brillouin zone of Si spans $2\pi/a = 11.5$ nm$^{-1}$. To cover the distance of ~9.2 nm$^{-1}$ from the $\Gamma$ point to the lowest energy near the X points ( ~1.1 eV, purple arrow and Gaussian momentum distribution in Figure 1b) in reciprocal space [26], a confinement of $\Delta r \sim 0.35$ nm is needed to permit direct photon momentum-enabled transitions near the indirect bandgap. For higher photon energies (transitions closer to the direct bandgap of Si) the required amount of confinement is less, for instance $\Delta r$ ~0.8 nm (633 nm, 1.96 eV, red arrow Figure 1b) and $\Delta r$ ~1.1 nm (532 nm, 2.33 eV). In all, the nanometer scale light confinement in this limit should translate into a broad momentum distribution, transforming otherwise indirect transitions across the Brillouin zone into direct, which can now be depicted as diagonal arrows in Figure 1b. An increase in the momentum of localized excitation fields has previously been suggested as an explanation for the enhancement of the interband transition rate in metals, as observed in two-photon electron emission experiments on roughened Ag films [27]. A different, but related effort has pointed out that the increased momentum associated with surface polaritons can be leveraged to enable indirect transitions in a variety of materials [28], and its application to indirect semiconductors has been insinuated [29]. For the case of bulk Si, we infer that the indirect-to-direct transformation enabled by photon confinement is accompanied by a significant boost of the transition probability, thus offering a mechanism for increasing the effective absorption coefficient of the semiconductor material across a broad spectral range, as illustrated in Figure 1c.

In this work, we experimentally demonstrate that such an effect is observed when light is confined on 1-3 nm asperities. We carry out measurements in three different experimental arrangements. In the first, as a proxy to absorption, we measure the heating of a silicon cantilever through Raman thermometry. We observe a dramatic increase in light absorption, which leads to melting of the silicon tip near ~1-2 nm structures and insignificant heating near >3 nm asperities, even though the latter structures sustain much larger plasmonic fields and thus a higher field enhancement. In a second configuration, we deposit nanometer-sized structures on a Si wafer and perform reflection measurements to demonstrate that it is the absorption that is enhanced across the full solar spectrum, down to silicon's band edge. In a final experiment, we show significant increase of the photocurrent in a Si photodiode when its surface is decorated with nanometric asperities. While these observations cannot be explained through plasmonic enhancement effects, we show that they follow the trends predicted by the mechanism of confined photon momentum.

RESULTS AND DISCUSSION

The first set of measurements is carried out with a tip-enhanced Raman scattering (TERS) microscope, equipped with a Si AFM cantilever oscillating in semi-contact mode over a planar substrate. The tip is illuminated with a focused, linearly polarized beam derived from a continuous laser at 633 nm (Fig. 2a and 2b). We record the Raman response of crystalline Si at 521 cm$^{-1}$, and monitor the evolution of the line intensity, spectral shift and width. The calibrated temperature dependence of this Raman mode makes it possible to estimate the absolute lattice temperature (see *Supplementary Information, Part 3*) [30,31]. Figure 2c shows the evolution of the Si Raman spectrum as a function of the incident laser intensity as the tip is positioned over a bare glass substrate. We observe a small blue-shift of the line as the laser intensity is raised, indicating heating. Under 5 MW/cm$^2$ irradiation levels, the line shifts by 1.5 cm$^{-1}$ (inset of Fig. 2c), corresponding to a temperature rise of $\Delta T$~70 K above ambient temperature (see *Supplementary Information* and Supplementary Figure SF2). Although these temperatures are well below the melting point of Si, we confirm through post-experiment SEM images that the tip retains its structural integrity during these Raman measurements (Figure 2f).

We observe a strikingly different behavior when the tip is placed over a 50 nm Au film, evaporated on a glass substrate (Fig. 2b). The smooth gold surface is characterized by height variations, with a mean value of $h_0$=1.7 nm and a standard deviation of 0.55 nm (Figure 2f and 2k). At irradiation intensities above 4 MW/cm$^2$, the spectrum splits into two distinct components, corresponding to contributions from the hot apex at wavenumbers below 500 cm$^{-1}$ and from the cold shaft at 521 cm$^{-1}$, which both occupy the collection volume of the objective lens (inset Figure 2d). At these laser intensities, we observe a 30 cm$^{-1}$ line shift, indicating a tip temperature rise in excess of $\Delta T$=1180 K (Figure 1d). As revealed through Raman imaging (see *Supplementary Information, Part 4*), the "hot" part extends about ~300 nm out from the apex (mapped at 480 cm$^{-1}$) into the Si material, whereas the colder part of the tip in the detection volume remains at ambient temperature (mapped at 521 cm$^{-1}$). The temperature near the apex exceeds the Si melting point (1683 K), conditions under which the tip melts and the apex pinches off, as verified by post-experiment SEM images (Figures 2g and 2h). In all tip-melting experiments, the pinch-off consistently happens at a distance of 150-250 nm from the apex. Remarkably, and as substantiated by the simulations below, such melting behavior would only be expected if input fluxes were ~25x higher than used in the current experiment (see *Supplementary Information Part 5* and Figure SF4).

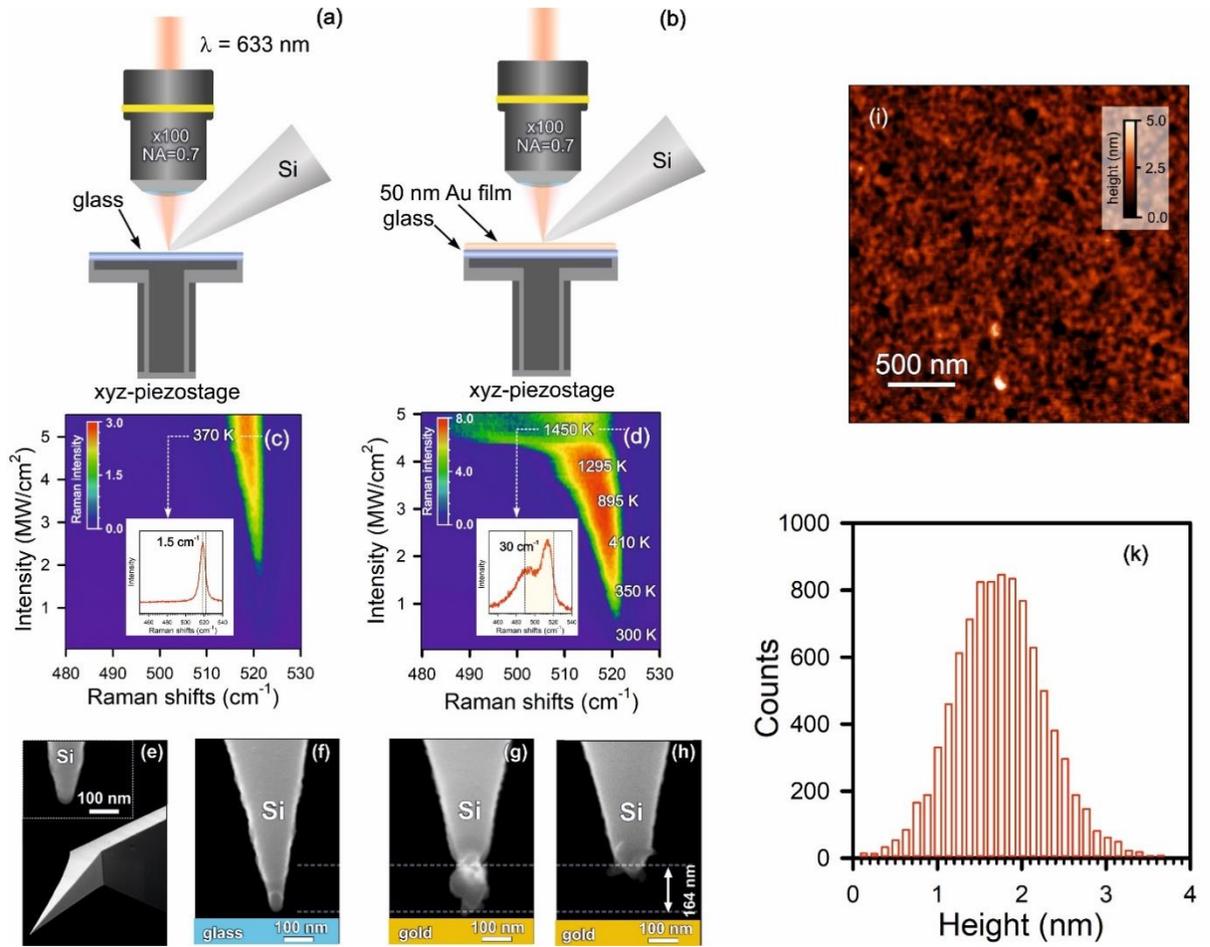

**Figure 2.** Experimental configuration when a Si AFM cantilever is placed over (a) bare glass and (b) a 50 nm gold film. Panels (c) and (d) show Si Raman spectra as a function of laser intensity for either bare glass or a smooth gold film. The insets indicate Raman spectra along the dashed straight line. (e-h) SEM images of an intact AFM cantilever with the tip apex before (inset) and after illumination for different substrates. (i) AFM image of the 50 nm thick Au film and its (k) surface roughness height distribution histogram centered at $h_0$=1.7 nm ($\sigma$=0.55 nm).

The experiments suggest an enhanced optical absorption in silicon when the tip is placed over a gold surface with nanoscale surface roughness, while the effect is absent on bare glass. Enhanced absorption can be anticipated when the incident field is locally enhanced at Au structures. However, the measured surface roughness of $h_0 = 1.7 \pm 0.6$ nm (Figure 2k) is well below the spatial scale needed to support strong localized surface plasmon resonances [32].

The heat dynamics in solids at the nanoscale can be rather complex, as diverse mechanisms can play a role (see *Supplementary Information Part 5*) [33]. To obtain reliable estimates of the tip temperature, we carry out FDTD/FEM simulations. The simulations show that, for a hemispherical Au protrusion of 2 nm, the penetration depth of the near-field into the Si tip apex is 2-3 nm (Supplementary Figure SF4) and the temperature increase at the tip apex is $\Delta T \sim 25$ K under the highest illumination doses used in the experiment (Supplementary Figure SF4e). The simulations also show that the temperature change of the Au substrate is much smaller, as would be expected based on its higher thermal conductivity. We note that the simulated temperature profile of the tip shows a plateau that stretches out to $z = 150$ nm, which becomes especially evident at higher illumination doses (yellow areas, Supplementary Figures SF4e and SF4f). This geometry-dependent thermal bottle-neck is likely due to the decrease of thermal conductivity $\kappa$ with temperature [34], and the size-dependent effect [30]. The simulations align closely with the experimental findings (Figure 1g and h), showing tip detachment at $z \sim 165$ nm. It is also clear that to initiate melting, a significant increase (>20 fold) of the maximum optical flux used in the experiment would be necessary, as illustrated in Figure SF4f. Both experimental data and simulations converge on the conclusion that *neither plasmonic effects nor geometry-induced field enhancement can account for the observed optical heating of the Si tip apex*, and that thus a different mechanism must be at play to explain the results.

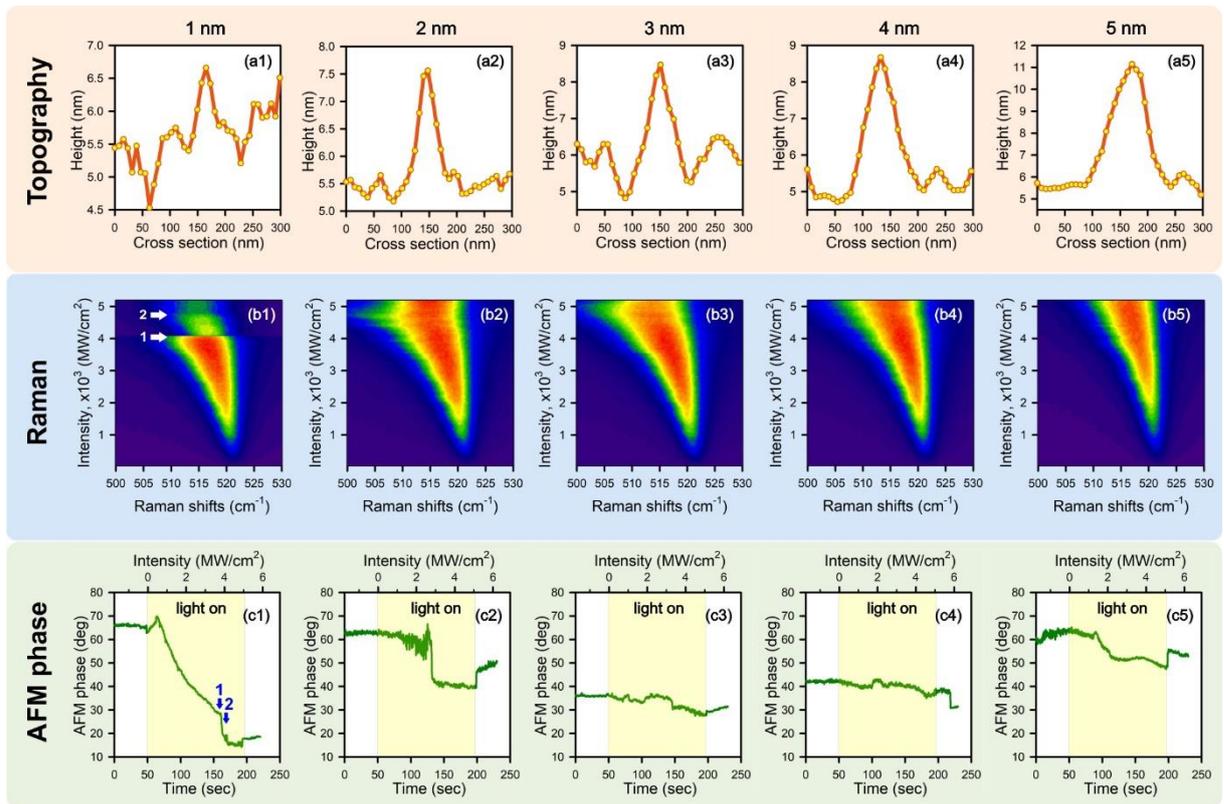

**Figure 3.** Sensing of tip apex temperature through Raman thermometry and cantilever phase evolution. (a1-a5) AFM topography cross-section of preliminary identified gold structures. (b1-b5) Raman spectrum evolution and (c1-c5) phase dynamic as a functions of input light intensity. The data clearly show that Si cantilever temperature is inversely proportional to structure size.

To further confirm the increased absorption in Si under photon confinement conditions, we perform measurements when the Si tip is placed over Au structures of a well-defined size. For this purpose, the film is analyzed by AFM imaging to identify individual and isolated Au structures within the 1-5 nm range (Figure 3a1-3a5). We record both the temperature dependence of the Raman spectrum (Figure 3b1-3b5) as well as the cantilever phase φ (Figure 3c1-3c5), which serves as another probe that is sensitive to changes in the tip-sample interaction. The phase is especially sensitive to the thermal expansion/contraction of the tip apex or the sample, which is particularly extreme during melting and deformation (see *Supplementary Information Part 5*).

We observe an inverse relationship between particle size and optical heating, as revealed by both Raman (Figures 3b2-3b5) and phase measurements (Figures 3c1-3c5). For a 1 nm Au structure, the Raman spectra indicate significant optical heating ($\Delta T$>1500 K), corroborated by the dramatic changes in the cantilever phase, both pointing to the melting of the tip. On the other hand, when the tip is brought into close proximity to structures larger than 3 nm, the temperature change is much smaller ($\Delta T$~200 K for 5 nm) and the tip apex remains fully intact. Note that because of irreversible changes in tip morphology, we have used different AFM cantilevers for different experiments. The observed temperature trends are reproducible with every cantilever used and over each gold structure identified. For example, as seen from Figure 3b1-b5, below a laser intensity of 3 MW/cm$^2$ the temperature change is fully reversible (Supplementary Figure SF5) and the tip apex remains unaffected. These experiments reveal that the optical absorption in Si grows much stronger when the size of the light-trapping nanostructures is decreased from 5 nm to 1 nm.

As a control experiment, we perform similar measurements with Si AFM cantilevers that are coated with a 30 nm Pt/Ir film. Using the cantilever phase as the readout, we observe a vanishingly small variation upon increasing the laser intensity, indicating insufficient optical heating and no structural damage of the tip or the sample (Supplementary Figure SF6). This observation underlines that close proximity of bare Si to a nanometer-size structure is required for the enhanced optical absorption in Si.

The anomalous heating of bare Si near the nm structure, the efficiency of which is inversely proportional to particle size, cannot be explained through field enhancement in the context of indirect optical transitions in Si. Note that this trend also excludes a prominent role for propagating surface plasmon polaritons at the gold/air surface, as their excitation efficiency at asperities is expected to grow with the size of the nanostructure, which is opposite of the heating trend observed. On the other hand, the highly localized optical fields at the apex of nanostructures in the range $\Delta r \sim 1 - 5$ nm can provide the necessary boost to the momentum of the photon $p \sim \pi\hbar/\Delta r$ to induce direct optical transitions (Fig. 1b). This mechanism of increased photon momentum is consistent with the observed inverse dependence of the absorption enhancement on the size of the nanostructure.

The altered absorption rates in silicon near nanometer-sized structures are also predicted by a simple rate analysis. The full effect of absorption enhancement has two contributions, namely the increase of photon momentum by 2-3 orders of magnitude and the Purcell effect. First, we assume a photon field of the form $E(r,t) = E_0 \hat{e} \cdot \mathcal{E}(r) e^{-i\omega t}$,[35] where $\omega$ is the angular frequency, $\hat{e}$ is the polarization state, $E_0 = \sqrt{\hbar\omega/(2V_0\varepsilon_0)}$, and $\mathcal{E}(r)$ is the spatial mode function. Fermi's golden rule for a transition from a given Bloch state $|v,k\rangle$ with crystal momentum vector $k$ in the valence band to all possible final states $|c,k'\rangle$ in the conduction band can then be written as:

$$W_{vc}(k) = \sum_{k'} \frac{2\pi}{\hbar} \left(\frac{e}{\omega m}\right)^2 \left(\frac{\hbar\omega}{2V_0\varepsilon_0}\right) |\langle c,k'|\mathcal{E}(r)\hat{e}\cdot\hat{p}|v,k\rangle|^2 \delta[\hbar\omega - E_{vc}(k,k')] \quad (1),$$

where $V_0 = q_0^{-3}$ is the mode volume of a photon with wave vector $q_0$ in free space, $E_{vc}(k,k')$ is the energy difference between band states, and $\hat{p}$ is the momentum operator. In the case of a free space photon, $\mathcal{E}(r) = e^{iq_0 r}$, thus the transition rate is found as:

$$W_{vc}(k) \propto \frac{1}{V_0} |\mu_{cv}|^2 \delta[\hbar\omega - E_{vc}(k, k+q_0)] \quad (2),$$

where $|\mu_{cv}|^2$ is the transition dipole moment. Since $q_0 \ll k$, only transitions for which $k' \approx k$ are optically allowed, which translates into vertical transitions. For the confined photon, we may assume the spatial mode function $\mathcal{E}(r)$ as a stationary Gaussian distribution with standard deviation $\sigma_r$:

$$\mathcal{E}(r) = e^{-\frac{r^2}{(2\sigma_r)^2}} = \frac{1}{2\pi^2}\int G(q) e^{iqr} d^3q \quad (3)$$

which is written here in terms of a spherically-symmetric, three-dimensional Fourier transformation of the momentum distribution function $G(q)$. In this scenario, Equation (1) yields for the transition rate:

$$W_{vc}(k) \propto \frac{1}{V_e}|\mu_{cv}|^2 \int d^3q \, e^{-\frac{q^2}{(2\sigma_q)^2}} \delta[\hbar\omega - E_{vc}(k, k+q)] \quad (4)$$

where $\sigma_q$ is the standard deviation of the photon wave vector distribution, and $V_e = \bar{\varepsilon}V$ is the effective mode volume with $\bar{\varepsilon} = \langle \mathcal{E}|\varepsilon|\mathcal{E}\rangle$ the expectation value of the local dielectric. In comparing Equations (2) and (4), it can be seen that this simple model predicts an enhanced absorption rate when the photon is confined to the nanometer scale. First, Equation (4) shows that transitions between different momentum states of the material are now allowed, weighted by the Gaussian distribution of photon momenta $\hbar q$. The transition further benefits from the integration over all $q$ made available by photon confinement, thus accelerating its rate relative to the free space result in Equation (2). Second, the local intensity is enhanced by the mode confinement $V_0/V_e$, where $V_0 = k_0^{-3}$ is the free space volume of the photon. However, our FDTD simulation shows that the field enhancement due to mode confinement is a relatively small effect, likely caused by a significant increase of $\bar{\varepsilon}$ in case of the nm-confined photon, thus reducing the $V_0/V_e$ enhancement effect. In short, field enhancement cannot explain the increased absorption on its own (see *Supplementary Information Part 7*), identifying the broadened photon-momentum distribution as the prime mechanism for the observed effect.

We expect that the spectral profile of the increased optical absorption due to sub-10 nm photon confinement is not colored by the spectral shape of a plasmon resonance. To examine the predicted spectral response of photon momentum-enabled transitions, we carry out reflectance measurements on Si wafers covered with gold nanoparticles. Deposition of nanoscale structures on Si surfaces and within the crystal bulk have been actively pursued for enhancing the efficiency of solar cells [16,36] and photodetectors [37]. Previous work has aimed at increasing the density of optical states through local field enhancement, using larger (>20 nm) particles and structures. In contrast, here we use Au structures that are significantly smaller, namely in the 1-2 nm range, where the plasmon resonance is strongly damped through surface scattering [38]. We deposit Au nanoparticles of a particular size distribution to form single-layer films in direct contact with the Si surface (see *Supplementary Information Part 8*). Figure 4a shows AFM images and the height distribution for samples prepared with ~1.3 nm and ~2.2 nm particles. The reflection spectra are obtained with an integrating sphere in diffuse reflection mode (*see Methods*) and are shown in Figure 4b. The difference in the reflectance spectra, between the bare wafer and the covered wafer, shows an abrupt reduction of light reflection at the band edge, at 1100 nm, and a lower but otherwise flat response throughout the UV-IR. The reduced reflectance can be explained as direct absorption, or equivalently, as the diagonal optical transitions shown in Fig. 1b. Consistent with the mechanism of increased photon momentum, the effect is inversely proportional to particle size (Figure 4b). This interpretation implies that upon illumination of the decorated surface, photons become trapped at nanometric asperities with finite probability and introduce a noticeable effect on the surface's optical properties. Note that the reflectance spectra are devoid of any spectral signatures of plasmonic resonances, and that silicon's absorption coefficient is uniformly enhanced along most of the solar spectrum. A rough quantitative estimation of the magnitude of absorption enhancement is provided in *Supplementary Information Part 10*.

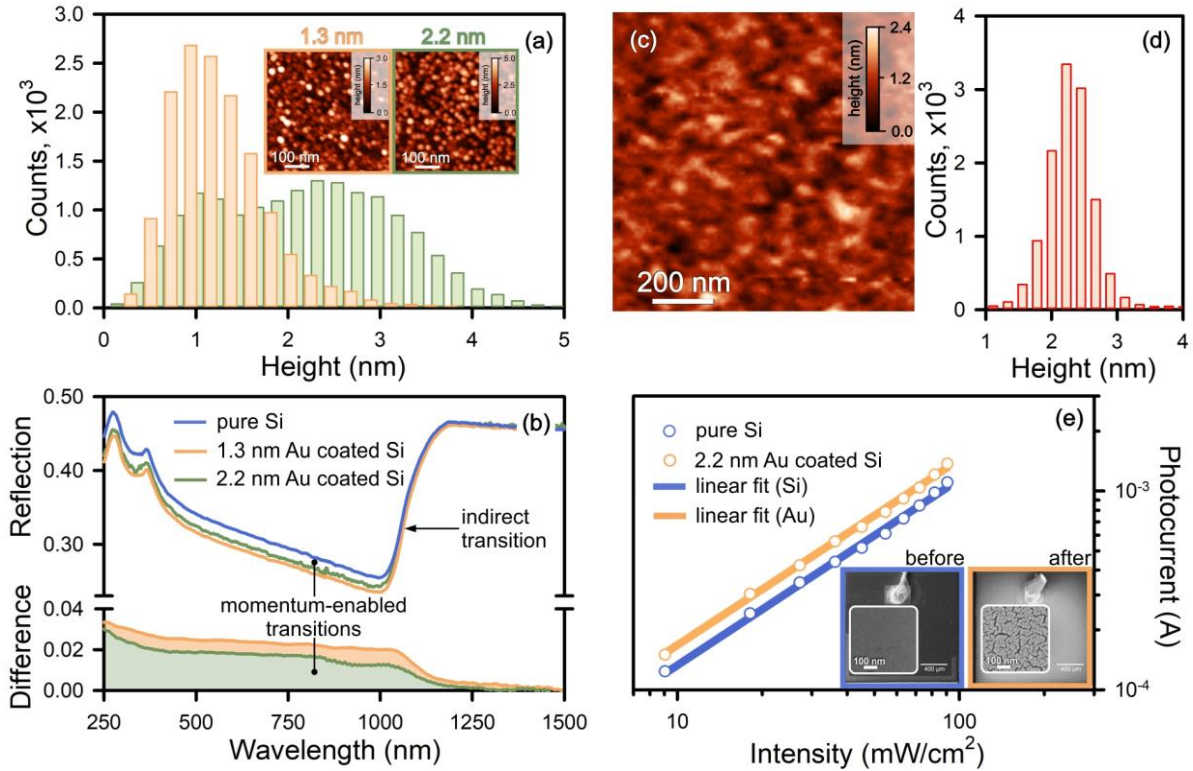

**Figure 4.** Experiments on Si wafers with deposited Au structures. (a) Height distribution histogram of single layer of Au nanoparticles of height ~1.3 nm and ~2.2 nm. (b) Visible/near-infrared reflection spectra of clean and coated Si wafers. Smaller particles provide broader momenta distribution, hence result in stronger and spectrally broader absorption enhancement. (c, d) AFM image and height distribution histogram of Au coated surface Si wafer after sputtering. (e) Photocurrent input power density dependence of Si photodiode with clean (blue) and coated (orange) photosensitive layer. The data shows a ~25% increase in detector responsivity for 2.2±0.5 nm Au structures.

Finally, in an arrangement more closely related to photo-detection and photovoltaic applications, using a different sputtering approach (see Methods and *Supplementary Information Part 11*, Figures SF11-SF14), we deposit Au structures on the top surface of the photosensitive layer of a conventional Si photodiode. A representative AFM image of the deposited Au structures, along with their height distribution, is shown in Figures 4c and 4d. Figure 4e shows the photocurrent detected from the clean (blue) and coated (orange) devices, measured as a function of the incident intensity of a 633 nm laser light source. The data shows an increase of detector responsivity from 0.33 A/W to 0.42 A/W, a nearly 25% increase in the amount of generated photoconduction electrons, despite the presence of Rayleigh scattering at larger Au structures and islands. While >50% of the light at 633 nm should be absorbed within the first 2.5 μm of the material[5], the observed detection enhancement corresponds to an absorption increase by *more than two orders* of magnitude within the first ~2 nm of the photosensitive layer. Similar to the tip heating experiments and the reflectance measurements on a decorated Si surface, the enhancement observed in the photocurrent experiment is in line with the hypothesis of increased absorption due to the widened momentum distribution of the confined photon.

CONCLUSIONS

Efforts to increase light absorption in silicon have hitherto focused primarily on trapping light in a given volume $V$ to increase its path length or field-enhancement mechanisms [11,16]. In the latter case, plasmonic antennas have been leveraged to redistribute the far-field intensity $I_0$ to a locally confined intensity $I=g^2I_0$, ($g$ is the field enhancement factor) in an attempt to increase the absorbed power within a particular volume $V$ to $P\sim \alpha_{indirect}VI=\alpha_{indirect}V g^2I_0$ [11,15,16]. In the present work, we achieve absorption enhancement via a different route, namely by the anomalous optical properties of silicon that are manifested when the indirect semiconductor is put in close proximity to structures of <3 nm in size. First, our TERS experiments with Si tips reveal an unusual optical heating effect at the nanoscale. Second, we have observed a similar enhanced optical response in reflection measurements on Si wafers decorated with nm-sized gold structures and in photocurrent measurements with Au-coated Si photodiodes. These phenomena cannot be explained through surface plasmon or geometry-dependent field enhancement mechanisms, as these are negligible for structures of this size. In addition, the absorption increase is inversely proportional to the size of the structures (Figure 3 and Figure 4), which is opposite to the simulated field-enhancement effects.

Our experiments emphasize that for light confined to nanometric scales its subsequent interaction with matter differs substantially from that enabled by plane waves. Confinement provides a boost of photon momentum by two orders of magnitude that opens up optical transitions forbidden for plane waves, effectively changing the material's absorption property. Whereas the light is confined, the semiconducting material is not necessarily altered, thus retaining its electronic structure as is the case in the experiments on bulk Si presented here. Nanometer confinement of the photon represents an alternative route to increasing the absorbed power $P= \alpha_{direct}VI$ through the effective change of the absorption coefficient $\alpha_{direct}$ of the medium. This enhancement is not limited to the momentum matching along the $\Gamma$ to X path in $k$-space, as shown for simplicity in Figure 1b. For a given photon energy, a broad momentum distribution enables transitions from multiple points and in any direction across the Brillouin zone, as long as available energies and momenta are conserved.

In contrast to the relatively narrow spectral width of plasmonic or geometry-governed resonances, the mechanism of increased photon momentum offers enhancement of the optical absorption from the ultraviolet to the near-infrared, turning an indirect semiconductor like silicon into a direct broadband absorber that resembles a black body absorber across 3 spectral octaves. The experimental results presented here provide an interesting opportunity to reconsider the role of photon momentum in light-matter interactions. With current advances in semiconductor fabrication techniques approaching a resolution of sub-1.5 nm [39], the effect has the potential to strongly impact not only optical spectroscopies but also photosensing and light-energy conversion technologies.

**METHODS**

**FDTD/FEM calculation**

3D simulation of optical absorption of a cone-shaped Si tip under cw focused illumination was performed by using an Ansys/Lumerical FDTD solver. A mesh overlayer of 0.1 nm was utilized around the Au bump and the Si tip apex and a rougher 1 nm mesh for the rest of the structure. Perfectly matching layers were used as boundary conditions for three directions. The optical and thermal properties of Si and Au were imported from the Ansys/Lumerical material database. The Si tip apex was exposed to a 632.8 nm focused laser light (NA= 0.7) with the intensity of 5 MW/cm$^2$. The temperature profile was calculated through an Ansys/Lumerical FEM solver in the steady state regime. The thermal conductivity of all constituents is assumed to be temperature-independent. The boundary condition of $T = 300$ K was set at the $z_{min} = -3500$ nm of the 20×20×5 μm$^3$ simulation region.

**Atomic force microscopy**

The multimode scanning probe microscope Prima (NT-MDT) was utilized for visualizing a topography of intrinsic Si wafer with sputtered Au nanoparticles. AFM cantilever (VIT_P) was made of antimony-doped single crystal silicon (n-type, 0.01-0.025 Ohm-cm). The tip height is 14-16 μm, the tip curvature radius is 30 nm, the resonant frequency was 300 kHz.

**Scanning electron microscopy**

The elemental composition, and morphology of the samples were studied by the Auriga Crossbeam Workstation (Carl Zeiss AG, Oberkochen, Germany), equipped with an INCA X-Max silicon drift detector (Oxford Instruments, Abingdon, UK) for energy dispersive X-ray microanalysis. For elemental analysis of Si diodes and wafers, an acceleration voltage of 5keV, an analytical working distance of 4 mm, and an electron probe current of 75 pA were used (see *Supplementary Information Part 7* for details).

**Light reflection measurement**

Total reflection spectroscopy on undoped 280 μm <100> Si wafer coated with Au nanoparticles were performed using Jasco V-670 absorption spectrometer equipped with 60 mm integrating sphere. All measurements performed at 200 nm – 2000 nm spectral range with 2 nm spectral resolution.

Specular reflection measurements on Si wafers sputtered with Au were performed using variable-angle spectroscopic ellipsometer (VASE by J.A. Woollam Co., Inc.) equipped with an auto-retarder. The reflection spectra were recorded across broad spectral range (250 nm to 2000 nm) with 400 μm beam spot size on the sample at 70° incident angle.

**Far- and near-field Raman spectroscopy and microscopy**

Raman spectra and maps were captured with a multi-purpose analytical instrument NTEGRA SPECTRA™ (NT-MDT) in both upright and inverted configuration. The confocal spectrometer was wavelength calibrated with a crystalline silicon (100) wafer by registering the first-order Raman band at 521 cm$^{-1}$. A sensitivity of the spectrometer was as high as ca. 2500 photon counts per 0.1 s provided that we used a 100× objective (N.A.=0.7), an exit slit (pinhole) of 100 µm and a linearly polarized light with the wavelength of 632.8 nm and the power at the sample of 10 mW. No signal amplification regimes of a Newton EMCCD camera (ANDOR) was used.

**Magnetron sputtering**

The surface of silicon wafers and diodes was coated with a gold (99.999% purity) nanolayer by using a sputter coater, Quorum Q150R ES Plus (Quorum Tech), for negative glow discharge at an applied current of 20 mA and sputter vacuum value of 10$^{-4}$ mbar. See *Supplementary Infromation Part 7* for details.

**ACKNOWLEDGEMENT**

D.A.F. and S.S.Kh. would like to thank Yulia Davydova and Natalia Bratkova for help and support. Authors thank Prof. Alexander Fishman, Prof. Sasha Chernyshev and William Harris for the fruitful discussions. S.S.Kh. thanks RSF grant No. 19-12-00066-P for support of experimental and theoretical studies of light absorption in confined silicon and Kazan Federal University Strategic Academic Leadership Program (PRIORITY-2030) for ellipsometry measurements. E.I.B. acknowledges support of FDTD/FEM calculations from Kazan Federal University grant FZSM-2022-0021. D.A.F and E.O.P. are thankful to Chan-Zuckerberg Initiative and grant 2023-321174 (5022) GB-1585590 for support in the development of Au/Si photodiode prototype. All authors acknowledge a technical support from NT-MDT BV (The Netherlands).